\documentclass[10pt]{article}
\usepackage{geometry} 
\usepackage{color}
\usepackage{graphicx}
\pdfoutput=1
\title{Shelf space strategy in long-tail markets}
\author{R. Alexander Bentley\\Anthropology Department, Durham University \\Durham DH1 3HN UK\\r.a.bentley@durham.ac.uk \and Paul Ormerod\\Institute of Advanced Study, Durham University \\ Volterra Consulting Ltd. \\ London SW14 8AE, UK\\pormerod@volterra.co.uk \and Mark E. Madsen\\Department of Anthropology, University of Washington \\Seattle, WA 98195 USA\\madsenm@u.washington.edu}
\begin{document}
\maketitle
\begin{abstract}
The Internet is known to have had a powerful impact on on-line retailer strategies in markets characterised by long-tail distribution of sales  \cite{Anderson_2004}.  Such retailers can exploit the long tail of the market, since they are effectively without physical limit on the number of choices on offer. Here we examine two extensions of this phenomenon.  First, we introduce turnover into the long-tail distribution of sales.  Although over any given period such as a week or a month, the distribution is right-skewed and often power law distributed, over time there is considerable turnover in the rankings of sales of individual products.  Second, we establish some initial results on the implications for shelf-space strategy of physical retailers in such markets.
\end{abstract}
%
%
\section{Introduction}
In his recent book,  \textit{The Long Tail}, Chris Anderson \cite{Anderson_2006}  qualitatively reviews how the Internet changes the dynamics of markets characterized by long-tailed distributions of sales, such that a few titles sell enormous amounts and most titles (the long tail) sell very little (Figure 1).  Recognized at least since the early 20th century \cite{Pareto_1907, Mandelbrot_1963, Zipf_1949} long-tailed distributions (and, specifically, power law distributions) in economics and society have been an exceedingly popular subject in the last 15 years  \cite{Borland_2002, Barabasi_Bonabeau_2003, Bentley_etal_2004, Bettencourt_etal_2007, Friedrich_etal_2000, Gabaix_etal_2006, McCauley_etal_2003, Mantegna_Stanley_1996, Newman_2005, Newman_etal_2006, Ormerod_2006}.
\\ \indent Although power laws have become a well-worn subject, with multiple potential causes recognized \cite{Newman_2005}, Anderson \cite{Anderson_2004, Anderson_2006} usefully identified a profound transition associated with the rise of Internet retailers, who can exploit the long tail of the market, as they are effectively without physical limit on the number of choices they can offer. A  retailer in a physical building, of course, cannot afford the space to stock the low-selling items beyond some point in the long tail (Figure 1). 
\\ \indent In a simplified sense, if a retailer has space for $y$ different items in the store, then a reasonable strategy is to stock the top $y$ best-selling items, as determined by market data.  In contrast, an Internet retailer \textit{can} sell items within the long tail, which can yield sales (area under the curve) comparable to those of the physical retailers selling the blockbusters at the top end (Figure 1).
%
%
\begin{figure}
\begin{center}
\includegraphics[width=5.5in]{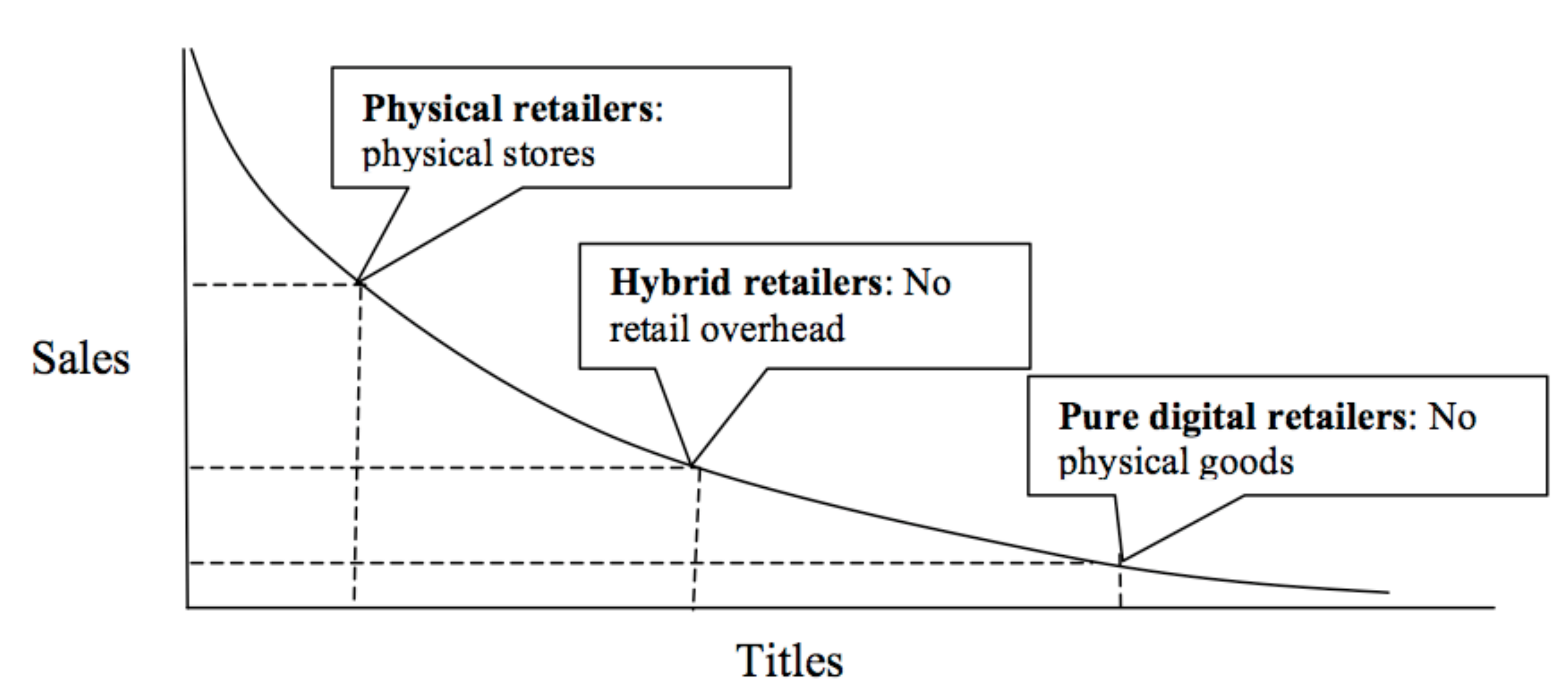}
\end{center}
\caption{Anderson's \cite{Anderson_2004, Anderson_2006} model of profit thresholds in an economy characterised by power law sales distributions, and physical retailers (real items from real shelves), digital retailers (digital goods from digital shelves) and hybrid retailers (physical goods from digital shelves).}
\end{figure}
\\ \indent Of course, in practice a retailer may want to follow alternative strategies from the one of stocking the top $y$ products.  Profit margins, for example, might be higher on certain products with low sales, the retailer may feel that specializing in a `niche' in the particular market may increase the chances of survival, and so on.  But for a retailer selling into the mass market, a good initial strategy to consider is one of selling the top $y$ items.  It is this strategy which we analyse and for which we establish initial results.
\\ \indent A crucial factor to take into account is the turnover in the sales (and hence the rankings) of individual products.  The distribution in Figure 1 is of course a stylized representation of the  power law distribution of sales at a \textit{given point in time}.  Over time, the relative popularity of the sales of the individual products will change.  So although  the distribution of sales may look very similar over time, taking snapshots of it at different points in time, the positions of the individual items within it will vary.  Indeed, in most fashion markets new items will constantly enter the market.
\\ \indent We consider a fashion-based model of consumer choice which is capable of generating power law distributions of sales such as are observed in practice, and which takes into account both turnover over time in the relative sales of a given set of items, and innovation in the sense that entirely new items become offered for sale.  The model has antecedents in the neutral model of population dynamics in biology \cite{Gillespie_2004, Nowak_2006}, but its relevance to long-tail consumer markets has been demonstrated by studies that show that the model is capable of explaining both the distribution of outcomes and the turnover over time in, for example, pop music, first names and dog breeds in the United States \cite{Hahn_Bentley_2003, Herzog_etal_2004, Bentley_etal_2007}.
\section{A fashion based model of consumer choice}
Standard consumer choice theory in economics assumes atomised individuals exercise choice in an attempt to maximise utility subject to a budget constraint.  In this approach, given an individualÕs tastes and preferences, decisions are taken on the basis of the attributes of the various products, such as price and quality.  
\\ \indent In recent decades, the conventional theory has been extended to allow for factors such as the cost of gathering information \cite{Stigler_1961}, imperfections in the perception of information and limitations to consumersÕ cognitive powers in gathering and processing information \cite{Simon_1955}.  So decisions are not necessarily made in a fully rational way, but are nevertheless based on the (perceived) attributes of the products, without direct reference to the choices of others (which can affect choice indirectly by their effect on relative prices).
\\ \indent Almost 50 years ago, Tintner \cite{Tinter_1960} extended classical consumer choice theory by assuming that individual utility depended not merely on an individualÕs consumption, but on the consumption patterns of other people.  This extension introduced far greater indeterminacy into the signs of income and price elasticities than standard theory.
\\ \indent In general, however, economists have paid little attention to markets in which fashion is important \cite{Chai_etal_2007}; i.e., markets in which the decisions of others can affect directly the choices made by an individual. Social influences are generally only invoked for cases considered exceptional, such as `irrational' stock market bubbles or real estate crises. 
\\ \indent Interest in such markets has, however, been much greater outside of economics.  There is strong empirical evidence that in markets where the decisions by others strongly influence individual choice, products that are superior in terms of their attributes may do no better than ones which are worse \cite{Salganik_etal_2006, Colbaugh_Glass_2007, Watts_Dodds_2007}. Even charitable donations can be highly subject to fashion \cite {Schweitzer_Mach_2008}. Such markets are characterised by randomness and inherent unpredictability \cite{Salganik_etal_2006}. 
\\ \indent An important development has been models of `binary choice with externalities' \cite{Schelling_1973, Watts_2002, Watts_Dodds_2007}.  In many social and economic contexts, individuals are faced with a choice between two alternative actions, and their decision depends, at least in part, on the actions of other individuals \cite{Ormerod_Wiltshire_2008}.  An important feature of such systems is that they are `robust yet fragile'.  In other words, behaviour may remain stable for long periods of time and then suddenly exhibit a cascade in which behaviour changes on a large scale across the individuals within the system \cite{Watts_2002, Ormerod_Colbaugh_2006, Watts_Dodds_2007}.
\\ \indent Our model has similar characteristics, but its relevance is to phenomena such as popular culture, in which consumers face many choices rather than  a simple binary decision between two alternative courses of action. Almost by definition, `popular culture' reflects the effects of most people imitating those around them. As recently re-emphasised by a music downloading experiment \cite{Salganik_etal_2006}, copied trends and fashions are constantly changing, with future outcomes potentially irrational and nearly impossible to predict. Considering markets dominated by peer-to-peer influence, our approach is essentially to model the  copying by individuals of choices made by others when many different choices are on offer.
\\ \indent In markets in which many products are on offer, with little to distinguish them in terms of their `objective' attributes, a sensible behavioural choice rule for an individual to adopt is one of random copying.  We can think of such markets as being `pure' fashion markets:  an individual copies the choice of another individual at random.  So at any point in time, the probability of choosing any particular item depends upon its relative popularity.  However, a further key feature of such markets is that there is constant innovation, in the sense that new products are being introduced all the time.
\\ \indent We populate the model with $N$ individuals, as well as $x_0$ products initially from which to choose. The number, $x$, of products on offer is allowed to vary over time, whereas the number of individuals, $N$, is fixed. When each individual has made a choice, the initial conditions are set. We start with one product per individual, or $N = x_0$, but the model is robust with respect to this initial condition (because the number of different products on offer soon stablises, as we discuss below).  
\\ \indent The model then proceeds in a series of steps.  In each step, each individual makes one further choice.  An additional $\mu N$ products are introduced, where $\mu \ll 1$ represents the fraction of individuals choosing novel items, via the $\mu N$  individuals (selected at random) who each choose a new product; each individual buys only one product. The remaining $(1 - \mu)N$ individuals select from the $x_{t-1}$ products available in the previous step.  Each individual selects product $i$ with the probability equal to product $i$'s proportion of total sales in the previous period. In other words, (s)he chooses product $i$ with probability $s_i/N$, where $s_i$ denotes the sales of product $i$ in the previous period.
\\ \indent So the choice rules have two components.  In each period, $(1-\mu)N$ individuals use a rule which is identical to that of the well-known model of preferential attachment, which is known to yield a long tail with asymptotic form $p(s)\sim s^{-3}$ \cite{Barabasi_Albert_1999}.  The remaining $\mu N$ individuals ($\mu$ is typically small, so these are a small minority) each select a product which was not previously on offer.  This second rule is essential in order for the model to be able to account for the turnover which is observed in the power law rank-size distribution of sales \cite {Bentley_etal_2007}.
\\ \indent  The model has been shown to yield a simple inverse power-law distribution with exponential cutoff for $N\mu$ equal to or somewhat greater than one, but for $N\mu \leq 1$ it yields a winner-take-all distribution, \textit{i.e.}, the same choice is made by most individuals \cite{Evans_2007}. Figure 2 (left) shows the distributions produced by the model for  a range of $N\mu$ values.  The cumulative sales $S_i$ of the products on offer follow a distribution which is a power law function for small values of the innovation fraction
\begin{equation}
P(S_i)= \frac{1}{S_i^\alpha}
\end{equation}
%
%
\begin{figure}
\begin{center}
\includegraphics[width=2.8in]{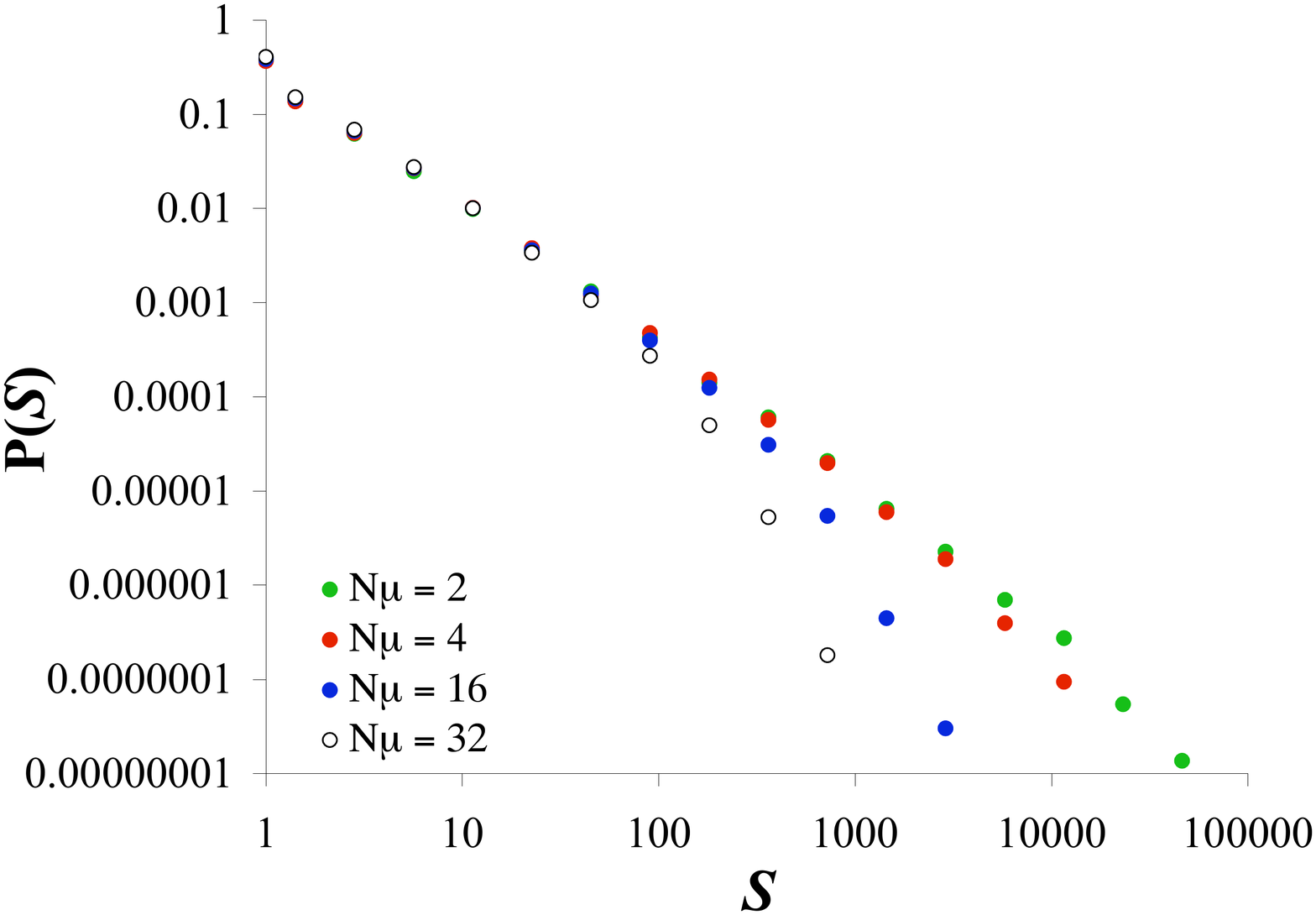}
\includegraphics[width=2.8in]{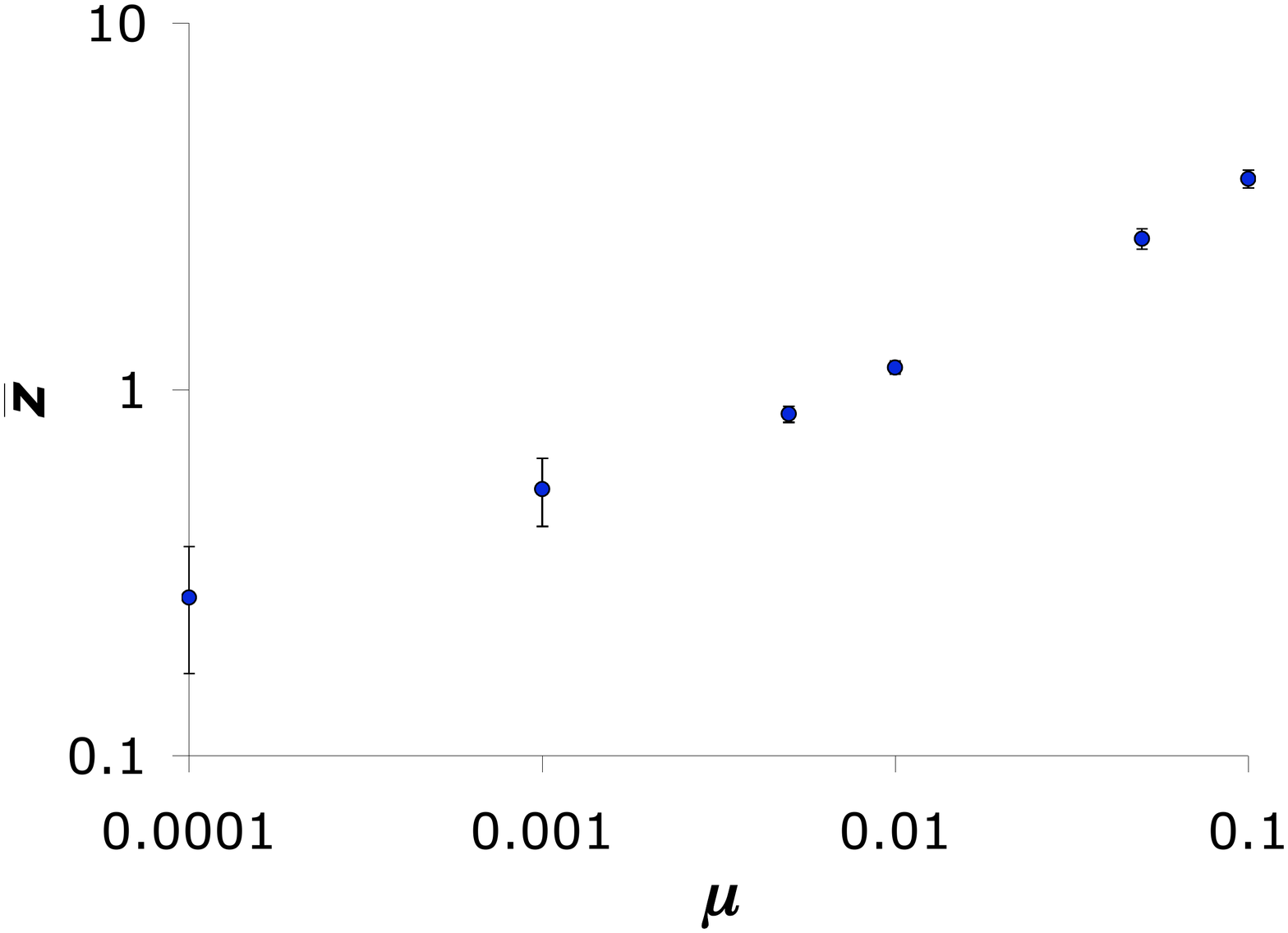}
\end{center}
\caption{Simulation results of the pure fashion model, showing double-log plots of: \textbf{(Left)} Probability $P(S)$ of accumulated sales $S$ for different values of $N\mu$. For $N\mu=2$ (green), the power-law exponent $\alpha$  is $1.54 \pm 0.01$, estimated by maximum likelihood \cite{Newman_2005}. \textbf{(Right)} Effects of innovation fraction $\mu$ on mean turnover rate $\bar{z}$ in the top 5 variants (600 runs of 1000 time steps; 10 runs averaged for each of 60 different combinations of $N$ and $\mu$).  The population size $N$ has only a small effect on turnover compared to $\mu$ \cite{Bentley_etal_2007}; the error bars show the standard deviation in runs from $N = 100$ to $N = 1,000$.  The best fit $(r^2 = 0.977)$ through the data is a function turnover $\sim \mu^{0.4}$.  As a simpler rule of thumb, a useful approximation is $z\approx y\sqrt{\mu}$ \cite{Bentley_etal_2007}.}
\end{figure}
\\ \indent There are many models that produce power law distributions, of course.  The most commonly proposed processes -- including preferential attachment and the Yule process  \cite{Newman_2005, Evans_2007} -- produce power laws via positive feedback in interactions between individual agents. While capturing much of the static distributions, strict preferential attachment models struggle to acount for how new entries can overtake old ones, and ways that the `rich' occasionally become extinct. What makes this model is the turnover which it produces within the consistuents of that distribution. 
\\ \indent Note that $x$, the number of products on offer at any one time, is never more than $N$, \textit{i.e.}, limited to those selected by $(1 - \mu)N$  individuals from the previous period, plus the new products adopted by $\mu N$  individuals in each period. A balance quickly arises between those new products being introduced, and those dropping out of the model because achieved zero sales in the previous period.  Given the rules of the model, there is no way of ever again choosing extinct products.  So most products over time become extinct, a well-known property not just of fashion markets, but of evolutionary systems more generally \cite{Ormerod_2006}. 
\\ \indent The model exhibits consistent average turnover rate on a `top y' list, as demonstrated by computer simulations \cite{Bentley_etal_2007}.  We define the turnover rate $z$ as the number of the $y$ best selling products that are new in any given period.  Products may drop out of the top $y$ because their sales are no longer large enough for them to qualify, and others may enter because their sales have increased.  The rate $z$ measures the net effect of these two.
\\ \indent Computer simulations of this model \cite{Hahn_Bentley_2003, Bentley_etal_2004, Bentley_etal_2007} have shown that the turnover rate is to a large extent independent of the population size $N$ (Figure 2, right), and its key determinants are the list size $y$ and the `novelty' parameter $\mu$, the fraction of previously unavailable products offered per time step \cite{Bentley_etal_2007}:
\begin{equation}
z\approx y\sqrt{\mu}
\end{equation}
\\ \indent The reason $N$ has little effect on turnover in the top $y$ is that, as $N$ increases, the greater number of newly-invented variants `trying' to reach the top $y$ is effectively counterbalanced by the increasing distance needed to reach it, textit{i.e.}, the lengthening tail of the rank-frequency distribution.
\\ \indent The modelled turnover fits evidence for steady turnover on real-world pop charts, independent of population changes, for pop music, first names, and dog breeds in 20th century United States \cite{Bentley_etal_2007}. The Billboard Top 200 album chart, for example, turned over at an average rate of 5.6\% per month (in terms of fraction of the 200 albums being replaced) for almost 30 years, from the 1950s to the 1980s \cite{Bentley_Maschner_1999, Bentley_etal_2007}.  
\section{Turnover and the long tail}
Consider a physical retailer that displays $y$ different items in a store.  The most sensible way to choose those $y$ items to display would be to simply take the top $y$ items from best-seller list.  If the top $y$ list were unchanging, there would be no need to change the inventory. Beststeller lists of course turn over constantly, however, so how does that affect the opimal inventory size $y$?  For a physical retailer,  there is a cost to changing the inventory as the composition of the top $y$ list changes.  For simplicity, let us assume for demonstration that the cost of changing the inventory is $B$, a fixed cost per new item to add to the inventory. 
\\ \indent Under our pure fashion model, the turnover rate is proportional to $y\sqrt{\mu}$, so the turnover cost is then $By\sqrt{\mu}$. For a power law distribution, sales of the top $y$ bestselling items can be approximated as $\sum_{i=1}^{y}1/{i^\alpha}$, and $A$ is the profit per item (we make the heroic assumption Ð for purposes of establishing initial results Ð that the profit per item is fixed). The retailer then seeks to maximise the difference between the profit obtained on the top $y$ best-selling items minus the cost of turnover:
\begin{equation}
A\sum_{i=1}^{y}\frac{1}{i^\alpha}-By\sqrt{\mu}
\end{equation}
This is maximized with respect to $y$ when, as $y$ is increased to the next higher rank, the additional term of  $A/y\alpha$ becomes less than $By\sqrt{\mu}$. This occurs where 
\begin{equation}
\frac{A}{y^{\alpha+1}}=B\sqrt{\mu},
\end{equation}
or
\begin{equation}
y=\frac{A}{B\sqrt{\mu}}^{\frac{1}{\alpha + 1}}.
\end{equation}
\\ \indent In other words, the optimal stock size in our hypothetical store is a function of the `innovation' fraction $\mu$, a sales/cost parameter $A/B$ and the power law exponent of the sales distribution, $\alpha$.
\\ \indent To take an example, the exponent $\alpha$ for book sales has been estimated to be $3.51 \pm 0.16$ \cite{Newman_2005}. As Figure 3 shows, the value of $y$ increases only modestly with $A/B$, the ratio of profit to turnover cost per item.  In fact, even if the profit $A$ is a hundred times the turnover cost $B$, the optimal inventory size $y$ is fewer than 20 items for a wide range of innovation parameter $\mu$. Although our model and hypothetical fixed cost assumptions are simple, the general form of the curves in Figure 3 suggest that the `Blockbuster' market could be very small indeed for physical retailers.
%
%
\begin{figure}
\begin{center}
\includegraphics[width=4.5in]{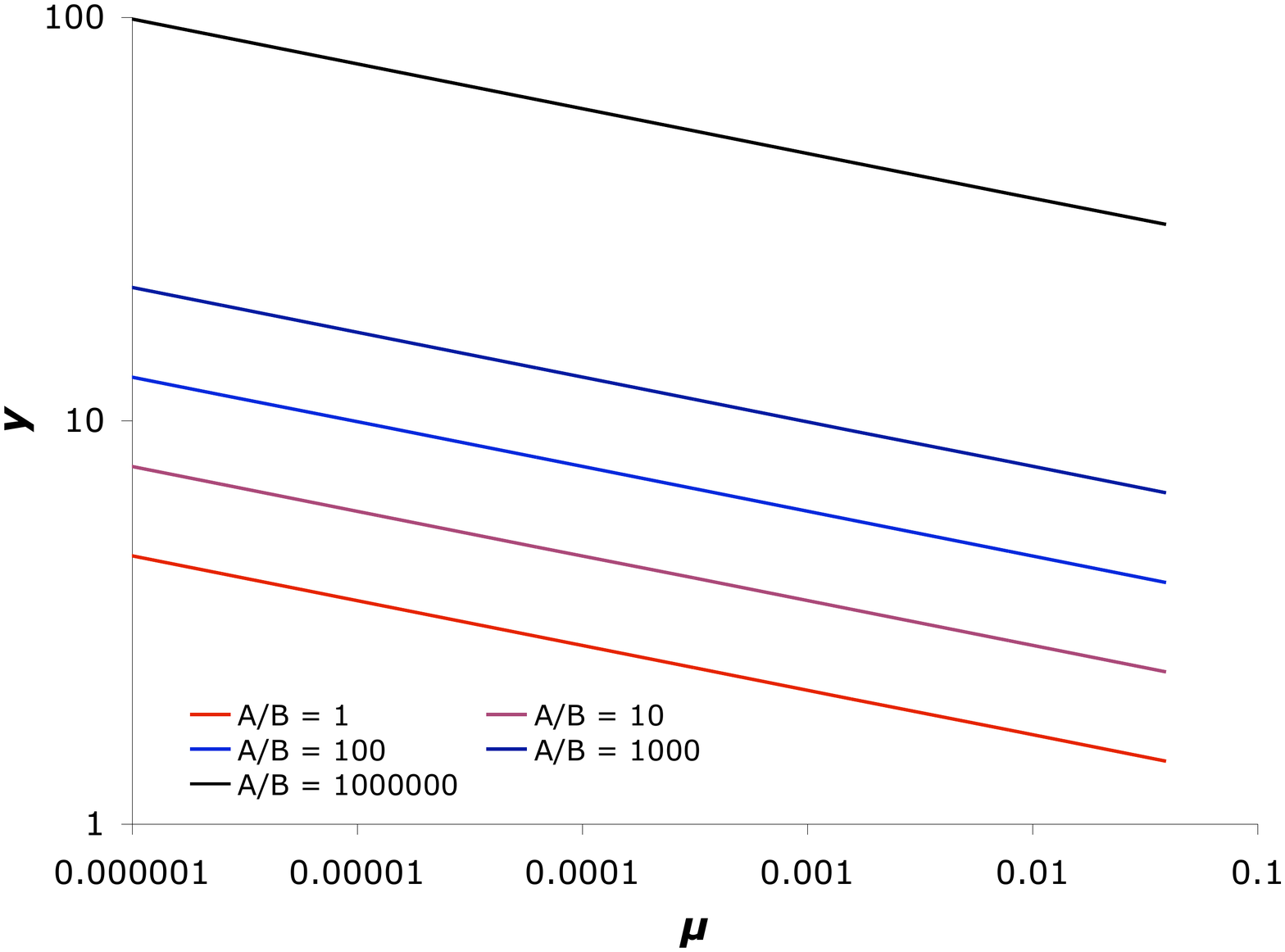}
\end{center}
\caption{The dependence of optimal inventory $y$ on the innovation parameter $\mu$, for a range of values of $A/B$.  Assumes $\alpha = 3.5$, as estimated for book sales \cite{Newman_2005}. Logarithmic axes.}
\end{figure}
\\ \indent This may seem extreme, but it is actually seen, for example, in book stores in busy train stations and airports, where the book stock often consists entirely of the top 10-20 fiction and non-fiction titles. For larger stores, it suggests it may make sense to track only the very top titles, while most of the rest of the titles, within the long tail, need not be rotated so often. 
\\ \indent There is another interesting implication for non-physical retailers, which can have extremely large profit/cost ratios $A/B$ per item, as Anderson \cite{Anderson_2004, Anderson_2006} pointed out, because $B$ is nearly zero.  The cost $B$ is not actually zero, however, so the ratio $A/B$ will never truly be infinite.  This is significant is because in the pure fashion model, the optimal $y$ increases only logarithmically with respect to $A/B$.  As Figure 3 shows, even if $A/B$ is one million the optimal stock is 100 or fewer, for a realistic range of $\mu$. In fact, if we assume $\mu$ were 0.1\%, then in order to bring $y$ to a million titles the profit/turnover ratio $A/B$ must  be increased to about $10^{30}$.  Although a very very rough calculation, this seems extreme even for digital media.
%
%
\section{Conclusion}
We conclude that, for digital retailers, the optimal inventory not only cannot be infinite in the literal sense, but may even be less than the millions of items offered by digital retailers like Amazon.com. Of course, more detailed estimates would depend on factors we have not accounted for in our simple model, such as the actual cost parameters for the company and the obvious prestige benefit of being a supplier that stocks virtually everything.  Nevertheless we generally concur with the recent caution \cite{Elberse_2008} that the long tail is by no means a guaranteed money-maker, as the sales proceeding further and further into the long tail become so small that he marginal cost of tracking them in rank order, even at a digital scale, might be optimised well before a million titles and certainly before infinite titles.  
\\ \indent We have characterized some aspects of change within the `Blockbusters' and `Long tail' portions of sales distributions which follow power laws.  In such markets, peer-to-peer influence is a crucial factor in consumer choice.  An individual is more likely to select a top selling item than one with low sales.
\\ \indent The basic behavioural choice rule followed by consumers is based upon the well-known model of preferential attachment.  Crucially, however, we modify this by allowing a small fraction of individuals in each period to choose items which have not previously been on offer. In this process Ð unlike strict preferential attachment Ð initially obscure new ideas can become highly popular by chance alone.
\\ \indent We can make predictions for the effects of population size and innovation on the turnover on bestseller lists.  In particular, the rate of innovation is more important than population; the turnover rate on a list of most popular variants should depend on the number of items on offer and the amount of innovation, but \textit{not} on the number of consumers \cite{Bentley_etal_2007}.  Increases in both population and/or innovation lead to increases in the increases the turnover in the Top y and therefore to a reduction in the number of titles to be treated as `blockbusters'. As innovation is increasingly encouraged by Internet blogs, youTube, music production software, and so on, the result should not only be that popular culture changes faster, but also that `Blockbuster' stores should shrink their inventory further and further.   
\\ \indent This model can provide further predictions into markets with long-tail distributions.  Usefully, these predictions derive from a minimum of parameters.  In future work, we hope to provide additional predictions of the  model, including characterizing lifespans on a best-seller list, and projected sales of an item on the top $y$ list (or at least the variance in those quantities) as function of its current rank.  This could provide a model for optimizing the number of each individual item ordered, given its current sales rank and the total number of different titles stocked.
\bibliographystyle{abbrv}

\end{document}